\documentclass[authoryear, a4paper]{article}\usepackage[]{graphicx}\usepackage[]{color}
\makeatletter
\def\maxwidth{ %
  \ifdim\Gin@nat@width>\linewidth
    \linewidth
  \else
    \Gin@nat@width
  \fi
}
\makeatother

\definecolor{fgcolor}{rgb}{0.345, 0.345, 0.345}

\usepackage{framed}
\makeatletter
 {\par\unskip\endMakeFramed%
 \at@end@of@kframe}
\makeatother

\definecolor{shadecolor}{rgb}{.97, .97, .97}
\definecolor{messagecolor}{rgb}{0, 0, 0}
\definecolor{warningcolor}{rgb}{1, 0, 1}
\definecolor{errorcolor}{rgb}{1, 0, 0}
\newenvironment{knitrout}{}{} 

\usepackage{alltt}
\usepackage[margin=3.4cm]{geometry}

\newtheorem{theorem}{\normalfont\scshape Theorem}
\newtheorem{algo}{\normalfont\scshape Algorithm}
\newtheorem{remark}{\normalfont\scshape Remark}

\usepackage{amsmath}

\usepackage{times}
\usepackage{bm}
\usepackage{natbib}
\usepackage[normalem]{ulem} 

\usepackage[plain,noend]{algorithm2e}

\makeatletter
\renewcommand{\algocf@captiontext}[2]{#1\algocf@typo. \AlCapFnt{}#2} 
\renewcommand{\AlTitleFnt}[1]{#1\unskip}
\def\@algocf@capt@plain{top}
\renewcommand{\algocf@makecaption}[2]{%
\addtolength{\hsize}{\algomargin}%
\sbox\@tempboxa{\algocf@captiontext{#1}{#2}}%
\ifdim\wd\@tempboxa >\hsize
\hskip .5\algomargin%
\parbox[t]{\hsize}{\algocf@captiontext{#1}{#2}}
\else%
\global\@minipagefalse%
\hbox to\hsize{\box\@tempboxa}
\fi%
\addtolength{\hsize}{-\algomargin}%
}
\makeatother

\usepackage{url}
\usepackage{amsmath}
\usepackage{amssymb}

\DeclareMathOperator{\E}{\mathnormal{E}}
\DeclareMathOperator{\Var}{var}
\DeclareMathOperator{\Cov}{cov}
\DeclareMathOperator{\Corr}{cor}
\DeclareMathOperator{\NegBin}{NBin}

\DeclareMathOperator{\diago}{diag}
\DeclareMathOperator{\vect}{vect}

\DeclareMathOperator{\AIC}{\textsc{aic}}
\DeclareMathOperator{\MPAR}{\textsc{mpar}}
\DeclareMathOperator{\PAR}{\textsc{par}}
\DeclareMathOperator{\hhh}{\textsc{hhh}}
\DeclareMathOperator{\varfct}{v}

\newcommand{\condmean}{\lambda}
\newcommand{\statmean}{\mu}
\newcommand{\lagparam}{\phi}

\newcommand{\given}{\,\mid\,} 
\usepackage{natbib}
\IfFileExists{upquote.sty}{\usepackage{upquote}}{}
\begin{document}


\title{Periodically stationary multivariate autoregressive models}
\author{Johannes Bracher and Leonhard Held}
\maketitle

\begin{center}
Epidemiology, Biostatistics and Prevention Institute, University of Zurich, Hirschengraben 84,\\ 8001 Zurich, Switzerland\\

\medskip

\texttt{johannes.bracher@uzh.ch \hspace{10mm} \texttt{leonhard.held@uzh.ch}}
\end{center}


\begin{abstract}
A class of multivariate periodic autoregressive models is proposed where coupling between time series is achieved through linear mean functions. Various response distributions with quadratic mean-variance relationships fit into the framework, including the negative binomial, gamma and Gaussian distributions. We develop an iterative algorithm to obtain unconditional means, variances and auto-/cross-covariances for models with higher order lags. Analytical solutions are given for the univariate model with lag one and multivariate models with linear mean-variance relationship. A special case of the model class is an established framework for modelling multivariate time series of counts from routine surveillance of infectious diseases. We extend this model class to allow for distributed lags and apply it to a dataset on norovirus gastroenteritis in two German states. The availability of unconditional moments and auto/cross-correlations enhances model assessment and interpretation.
\end{abstract}

\maketitle


\section{Introduction}

Periodic autoregressive models \citep{Troutman1979} are a natural way to capture autocorrelation and seasonality in time series and are commonly used in disciplines like hydrology 
\citep{Tesfaye2006} and econometrics 
\citep{Bollerslev1996}. In infectious disease epidemiology they are often applied to time series of counts from routine surveillance systems for infectious diseases \citep{Morina2011, Corberan-Vallet2014}. Accounting for seasonality, a common feature of many infectious diseases \citep{Grassly2006}, is crucial in this context.  An established modelling framework is the so-called \textit{endemic-epidemic} or $\hhh$ class \citep{Held2005}, a multivariate time series model offering modelling strategies for disease spread in space \citep{Meyer2014} and across age groups \citep{Meyer2017a}. In \citet{Held2017} we recently proposed a recursive algorithm to compute predictive first and second moments as well as auto- and cross-correlations for use in long-term forecasting.

In this article we study their unconditional counterparts, thus providing useful insights on model properties and a theoretical foundation for the endemic--epidemic model. This goes along with a generalization to a wider class of multivariate periodic autoregressive models with coupling between time series via linear conditional means. Specifically, we include higher order lags and broaden the range of possible distributional assumptions. We provide analytical results for the univariate model with lag one and multivariate models with linear mean-variance relationship. The multivariate case with higher order lags and quadratic mean-variance relationship can be solved numerically. Knowledge of unconditional moments, auto- and cross-correlations can greatly facilitate model assessment and interpretation, especially in the multivariate case. We illustrate this in a case study on incidence of norovirus gastroenteritis in the German states of Bremen and Lower Saxony. We apply an endemic--epidemic model with distributed lags to the bivariate time series of case counts and use the proposed techniques to explore properties of the fitted model. The use of distributed lags improves the model fit considerably.

The article is structured as follows. We define the multivariate periodic autoregressive model in Section \ref{MPAR} and give an overview of related models from the literature. Periodic stationarity properties are addressed in Section \ref{sec:periodic_stationarity}. In Section \ref{sec:application} we present the endemic--epidemic model as a special case and apply it to our example data set. Section \ref{sec:conclusion} concludes with a brief outlook.

\section{The multivariate periodic autoregressive model}
\label{MPAR}

In its general form we define the multivariate periodic autoregressive model $\MPAR$($G, Q, R$) as a $G$-dimensional vector process $\{\mathbf{Y}_t = (Y_{1t}, \dots, Y_{Gt}); t \in \mathbb{N}\}$ with mean process
\begin{equation}
\bm{\lambda}_t = \bm{\nu}_t + \bm{\lagparam}_{1t} \mathbf{Y}_{t - 1} + \ldots + \bm{\lagparam}_{Qt} \mathbf{Y}_{t - Q} + \bm{\kappa}_{1t} \bm{\lambda}_{t - 1} + \ldots + \bm{\kappa}_{Rt} \bm{\lambda}_{t - R}\label{eq:model_equation_matrix}
\end{equation}
and conditional response model
\begin{equation}
Y_{gt}\given \mathcal{F}_{t - 1}  \sim F(\condmean_{gt}, \psi_{gt});\ \ Y_{gt} \bot Y_{g't} \given \mathcal{F}_{t - 1}.\label{eq:dist_ass}
\end{equation}
Here $F$ is a probability distribution parametrized by its mean $\lambda_{gt}$ and an additional parameter $\psi_{gt}$, usually a dispersion parameter. The $\sigma$-field $\mathcal{F}_{t - 1}$ is generated by $\mathbf{Y}_{t - 1}, \bm{\lambda}_{t - 1}, \mathbf{Y}_{t - 2}, \bm{\lambda}_{t - 2}, \dots$ and represents all the information available up to time $t - 1$. The parameters $\bm{\nu}_{t}, \bm{\psi}_{t} \in \mathbb{R}_{\geq 0}^G$ and $\bm{\lagparam}_{qt}, \bm{\kappa}_{rt} \in \mathbb{R}_{\geq 0}^{G\times G}$ are constrained to be non-negative and periodic, i.e.\ for period length $L$ we assume that $\bm{\nu}_t = \bm{\nu}_{t + L}, \bm{\phi}_{qt} = \bm{\phi}_{q, t + L}, \bm{\kappa}_{rt} = \bm{\kappa}_{r, t + L}, \bm{\psi}_{t} = \bm{\psi}_{t + L}$. The restriction to non-negative parameters can be omitted for certain specifications of $F$ (e.g.\ the normal distribution). However, in practice our focus is on distributions with positive support and so this restriction is necessary to ensure well-definedness. The model then only allows for positive auto- and cross-correlations. Log-linear models \citep{Fokianos2011} can avoid this restriction but are not addressed here.

As we will also see in the application discussed in Section \ref{sec:results} (Fig. \ref{fig:correlations}), the conditional independence assumption $Y_{gt} \bot Y_{g't} \given \mathcal{F}_{t - 1}$ does not imply the absence of unconditional auto- and cross-correlations between time series; these arise from the coupling of the mean structure. We motivate the assumption for time series of infectious disease counts in Section \ref{subsec:hhh4}.

While we do not require $F$ in \eqref{eq:dist_ass} to be from the exponential family, we restrict attention to distributions with a quadratic variance function \citep{Morris1982}
\begin{equation}
\Var(Y_{gt}\given \mathcal{F}_{t - 1}) = \varfct_{gt}(\lambda_{gt}) = a_{gt}\lambda_{gt}^2 + b_{gt} \lambda_{gt} + c_{gt}.\label{eq:abc}
\end{equation}
In the context of univariate AR(1) models, distributions with quadratic variance functions have been studied by \cite{Grunwald2000} who introduced a class of conditional linear autoregressive (\textsc{clar}) models. We extend this to the multivariate periodic case with higher order lags. The terms $a_{gt}, b_{gt}$ and $c_{gt}$ are functions of $\psi_{gt}$ and the distributional assumption for $F$. They play a central role in the context of second-order periodic stationarity and allow for a simple yet general display of important formulas. Table \ref{tab:qrs} provides them for six standard distributions, along with the re-parametrizations of these distributions in terms of their mean $\lambda_{gt}$ and a dispersion parameter $\psi_{gt}$. For more details on the Gaussian model we refer the reader to \citeauthor{Franses2004} (\citeyear{Franses2004}; Chapter 5). The univariate Poisson version of our model is discussed in \cite{Bentarzi2017}; this has been extended to Poisson mixture distributions in a recent report by \cite{Aknouche2017}, but without addressing unconditional moments.

\begin{table}[h]
\caption{Re-parametrization of six standard distributions}
\begin{center}
\def\arraystretch{1.3}
\begin{tabular}{l  c c  c c  c c c c}
& Par. 1 &Par. 2 & $\lambda_{gt}$ & $\psi_{gt}$ & $a_{gt}$ & $b_{gt}$ & $c_{gt}$ \\
\hline
neg. bin. & $r^*_{gt} > 0$ & $\pi^*_{gt} \in (0, 1)$ & $\frac{\pi^*_{gt}r^*_{gt}}{1 - \pi^*_{gt}}$ & ${r^*_{gt}}$ & $\frac{1}{\psi_{gt}}$ & 1 & 0\\
Poisson & $\lambda^*_{gt} > 0$ & -- & $\lambda^*_{gt}$ & -- & 0 & 1 & 0 \\
gamma & $\alpha^*_{gt} > 0$ & $\beta^*_{gt} > 0$ & $\frac{\alpha^*_{gt}}{\beta^*_{gt}}$ & $\alpha^*_{gt}$ & $\frac{1}{\psi_{gt}}$ & 0 & 0 \\ 
Gaussian & $\mu^*_{gt} \in \mathbb{R}$ & $\sigma^{*2}_{gt} > 0$ & $\mu^*_{gt}$ & $\sigma^{*2}_{gt}$ & 0 & 0 & $\psi_{gt}$ \\
Laplace & $\mu^*_{gt} \in \mathbb{R}$ & $\sigma^*_{gt} > 0$ & $\mu^*_{gt}$ & $\sigma^*_{gt}$ & 0 & 0 & $2\psi_{gt}^2$ \\
uniform & $\alpha^*_{gt} \in \mathbb{R}$ & $\beta^*_{gt} > \alpha^*_{gt}$ & $\frac{\alpha^*_{gt} + \beta^*_{gt}}{2}$ & $\frac{\beta^*_{gt} - \alpha^*_{gt}}{2}$ & 0 & 0 & $\frac{\psi_{gt}^2}{3}$
\end{tabular}
\end{center}
\label{tab:qrs}
\begin{footnotesize} The first two columns contain the classical parametrizations of the different distributions. To avoid confusion with the use of the symbols $\mu, \sigma^2, \alpha$ and $\beta$ in Sections \ref{sec:periodic_stationarity} and \ref{subsec:hhh4} we added asterisks $^*$ to these parameters. The third and fourth column contain a re-parametrization in terms of the mean $\lambda_{gt}$ and a dispersion parameter $\psi_{gt}$ as required in equation \eqref{eq:dist_ass}. The three remaining columns contain parameters $a_{gt}, b_{gt}$ and $c_{gt}$ to map the mean of the distribution to its variance via $\varfct_{gt}(\lambda_{gt})$; see equation \eqref{eq:abc}. In principle the beta and the binomial distribution also fall into the class of possible distributions, but we omitted them as somewhat cumbersome constraints on the parameter space are required to ensure well-definedness in these cases. Consider also Table 1 in \cite{Morris1982}.
\end{footnotesize}
\end{table}

Further related work can be found in the area of integer-valued autoregression. Important univariate time-homogeneous models include the autoregressive Poisson \citep{Ferland2006, Fokianos2009} and negative binomial \citep{Zhu2011} models. Various multivariate generalizations exist, e.g.\ \cite{Heinen2007} and a recent report by \cite{Doukhan2017}; see also \cite{Karlis2015} for an overview. A somewhat different approach are thinning operator based INAR (integer valued autoregressive) models, see \cite{Monteiro2010} for a periodic and \cite{Pedeli2013} for a multivariate model version. In all the multivariate approaches the focus is on relaxing the conditional independence assumption $Y_{gt} \bot Y_{g't} \mid \mathcal{F}_{t - 1}$ while our goal is to extend the model to higher order lags and periodicity.

\section{Periodic stationarity properties}
\label{sec:periodic_stationarity}

\subsection{Definitions and notation}

A $G$-dimensional vector process $\{\mathbf{Y}_t = (Y_{1t}, \dots, Y_{Gt}); t\in \mathbb{N}\}$ is called periodically stationary in the mean with period $L$ if there exist $\statmean_{gt} = \E(Y_{gt})$ such that $\mu_{gt} = \mu_{g, t + L}$. The process is called periodically stationary of second order if $\sigma^2_{gt} = \Var(Y_{gt});\ \sigma^2_{gt} = \sigma^2_{g, t+L}$ and $\gamma_{g'gt}(d) = \Cov(Y_{gt}, Y_{g', t - d});\ \gamma_{g'gt}(d) = \gamma_{g'g, t + L}(d)$ exist for any $d \in \mathbb{N}$ and $g, g' = 1, ..., G$.

In the following we study the model in three steps. First we address the $\MPAR(1, 1, 1)$ model for which analytical results are feasible. We then move to $\MPAR(G, Q, 0)$ models. Analytical solutions are available if the conditional mean-variance relationship of the model is linear; if it is quadratic we resort to numerical solutions based on a recursive relationship. The suggested methods then translate easily to the full $\MPAR(G, Q, R)$ model.

\subsection{Analytical results for the univariate model with $G = Q = R = 1$}
\label{subsec:analytical_results}

In this section we provide some analytical results for the simplest model version $\MPAR$(1, 1, 1), or more concisely $\PAR$(1, 1). For 
the Poisson case \citep{Bentarzi2017} these are already available elsewhere; in the case of time-constant parameters and $\kappa = 0$ they reduce to those given in \cite{Grunwald2000}. We drop the indices $g, q$ and $r$ and simplify equation \eqref{eq:model_equation_matrix} to
\begin{equation}
\E(Y_t \given Y_{t - 1}, \lambda_{t - 1}) = \lambda_t = \nu_t + \phi_t Y_{t - 1} + \kappa_t \lambda_{t - 1}.\label{eq:means_univ}
\end{equation}
The following results hold for all distributions in Table \ref{tab:qrs}; proofs can be found in the Appendix. To make notation easier we set $\prod_{i = j}^k x_i = 1$ for $j > k$ in equations \eqref{eq:theorem_means}, \eqref{eq:theorem_vars} and \eqref{eq:theorem_cor}.

\begin{theorem}
\label{theorem:means}
A $\PAR$(1, 1) process $\{Y_t; t \in \mathbb{N}\}$ is periodically stationary in the mean if $\prod_{m = 1}^L (\lagparam_{m} + \kappa_m) < 1$. The unconditional means are then
\begin{equation}
\mu_{t} = 
\frac{\sum_{i = 0}^{L - 1} \left(\nu_{t - i} \prod_{l = 0}^{i-1} (\lagparam_{t - l} + \kappa_{t - l})\right)}{1 - \prod_{m = 1}^L (\lagparam_{m} + \kappa_m)}.\label{eq:theorem_means}
\end{equation}
\end{theorem}
A $\PAR(1, 1)$ process can thus resemble a non-stationary process for some $t$, i.e.\ $(\phi_t + \kappa_t) > 1$, and as a whole still be periodically stationary. Especially when modelling case counts of seasonal infectious diseases this is a desirable property.

\begin{theorem}
\label{theorem:second_mom}
A $\PAR$(1, 1) process $\{Y_t; t \in \mathbb{N}\}$ is periodically stationary of second order if $\prod_{m = 1}^L h_m < 1$ with $h_m = \left\{(\phi_m + \kappa_m)^2 + \phi_m^2a_{m - 1}\right\}$.
The unconditional variances are then
\begin{align}
\sigma^2_t & = \varfct_t(\mu_t) + (a_t + 1) \Var(\lambda_t)\nonumber\\
& = \varfct_t(\mu_t) +  (a_t + 1) \ \frac{\sum_{i = 0}^{L - 1} \phi_{t - i}^2 \varfct_{t - i - 1}(\mu_{t - i - 1}) \prod_{l = 0}^{i - 1} h_{t - l}}{1 - \prod_{m = 1}^{L} h_{m}}.\label{eq:theorem_vars}
\end{align}
\end{theorem}
The unconditional variance $\sigma^2_t$ is thus a weighted sum of the (conditional) variance functions $\varfct_m$ evaluated at the respective unconditional means $\mu_m$. The condition for second-order stationarity is stricter than the one for stationarity in the mean if there is any $a_{gt} > 0$.  In this case we can construct models with finite means but infinite variances.

\begin{theorem}
\label{theorem:covcor}
Under the same conditions as in Theorem \ref{theorem:second_mom} unconditional $d$th order autocovariances
\begin{align}
\gamma_{t}(d) = \Cov(Y_t, Y_{t - d}) & = \left\{\prod_{i = 0}^{d - 2}(\phi_{t - i} + \kappa_{t - i})\right\} \left\{\phi_{t - d + 1} \sigma^2_{t - d} + \kappa_{t - d + 1} \frac{\sigma^2_{t - d} - \varfct_{t - d}(\mu_{t - d})}{a_{t - d} + 1}\right\}\label{eq:theorem_cor}
\end{align}
and autocorrelations $\rho_t(d) = \gamma_t(d)/(\sigma_t\sigma_{t - d})$ exist for $d = 1, 2, \dots$
\end{theorem}
Theorem \ref{theorem:covcor} implies that $\gamma_{t}(d) = (\phi_t + \kappa_t)\gamma_{t-1}(d - 1)$ for $d > 1$. For $\kappa_t \equiv 0$ we can obtain $\gamma_{t}(d) = \sigma_{t - d}^2\prod_{i = 0}^{d - 1} \phi_{t - i}$ and $\rho_t(d) = \prod_{i = 0}^{d - 1}\rho_{t - i}(1)$. This is the periodic equivalent of the exponentially decaying autocorrelation function often found in time-homogeneous autoregressive models.

For univariate models with linear conditional mean-variance relationship, notably the Poisson model, closed forms for unconditional moments are also feasible for $Q, R > 0$; see Remarks \ref{remark_mu}, \ref{remark_linear} and Theorem \ref{theorem:linear}. However, the formulas are not very enlightening and so we omit detailed discussion.

\subsection{Analytical and numerical solutions for the multivariate model with $Q > 0, R = 0$}
\label{subsec:algorithm_moments}

In practice we are more interested in the multivariate version of our model and would also like to include higher order lags. We first consider the case $Q > 0, R = 0$ which is also the type of model we use in our data example in Section \ref{sec:application}. To this end we introduce column vectors $\tilde{\mathbf{Y}}_{t} = (1, \mathbf{Y}_{t - Q + 1}^\top, \dots, \mathbf{Y}_{t}^\top)^\top \in \mathbb{R}^{1 + QG}$ which contain 1 as first entry followed by the stacked observations of $Q$ time periods leading up to and including $t$. The corresponding unconditional mean vectors are denoted by $\tilde{\bm{\statmean}}_{t} = \E(\tilde{\mathbf{Y}}_{t})$. The unconditional auto- and cross-covariance structure of the process is addressed in the form of matrices $\tilde{\mathbf{M}}_{t} = \E(\tilde{\mathbf{Y}}_{t}\tilde{\mathbf{Y}}_{t}^\top)$. To this end we arrange the model parameters into matrices
\begin{align*}
\tilde{\bm{\lagparam}}_{t} = \left(\begin{array}{ c c c c c }
1 & \mathbf{0}_{1 \times G} &  \multicolumn{3}{c}{\mathbf{0}_{1 \times (Q - 1)G}}\\
\mathbf{0}_{(Q - 1)G\times 1} \ \ & \mathbf{0}_{(Q - 1)G\times G} \ \ & \multicolumn{3}{c}{\mathbf{I}_{(Q - 1)G\times (Q - 1)G}} \\
\bm{\nu}_t & \bm{\lagparam}_{Qt} & \bm{\lagparam}_{Q - 1, t} & \cdots & \bm{\lagparam}_{1t}
\end{array}\right) \in \mathbb{R}_{\geq 0}^{(1+QG)\times(1 + QG)}.
\end{align*}
The dispersion parameters $a_{gt}, b_{gt}$ and $c_{gt}$ (Table \ref{tab:qrs}; equation \eqref{eq:abc}) are assembled in column vectors $\tilde{\mathbf{a}}_t = (\bm{0}_{1 + (Q - 1)G}^\top, a_{1t}, \dots, a_{Gt})^\top$, $\tilde{\mathbf{b}}_t = (\bm{0}_{1 + (Q - 1)G}^\top, b_{1t}, \dots, b_{Gt})^\top, \tilde{\mathbf{c}}_t = (\bm{0}_{1 + (Q - 1)G}^\top, c_{1t}, \dots, c_{Gt})^\top$ which each contain $1 + (Q - 1)G$ leading zeros.

\begin{remark}
In the case where the $\tilde{\bm{\statmean}}_{t}$ exists, $\tilde{\bm{\lagparam}}_{t}$ is constructed such that $\tilde{\bm{\statmean}}_{t} = \tilde{\bm{\lagparam}}_{t} \tilde{\bm{\statmean}}_{t - 1}$. Applying this relationship iteratively and setting $\tilde{\bm{\Phi}}_t = \tilde{\bm{\lagparam}}_{t} \tilde{\bm{\lagparam}}_{t - 1} \cdots \tilde{\bm{\lagparam}}_{t - L + 1}$ we get $\tilde{\bm{\Phi}}_t\tilde{\bm{\mu}}_t = \tilde{\bm{\mu}}_t$. Thus $\tilde{\bm{\mu}}_t$ is the real-valued non-zero eigenvector of the matrix $\tilde{\bm{\Phi}}_t$ which corresponds to an eigenvalue of 1, normalized such that its first element equals 1. The strategy of obtaining moments of periodic processes as the solution of eigenvalue problems has been suggested by \cite{Ula1997}.\label{remark_mu}
\end{remark}

\noindent We now establish a recursive relationship between the $\tilde{\mathbf{M}}_t$.

\begin{theorem} Provided that an $\MPAR(G, Q, 0)$ process is second-order periodically stationary, its unconditional first and second moments can be expressed recursively as \label{theorem:recursion}
\begin{align}
(\tilde{M}_{t})_{ij} & = \begin{cases}
\{(\tilde{a}_t)_i + 1\}(\ddot{M}_{t})_{ii} + (\tilde{b}_t)_i (\ddot{M}_{t})_{1i} + (\tilde{c}_t)_i & \text{ if } i = j\\
(\ddot{M}_{t})_{ij} & \text{ otherwise}
\end{cases}\label{eq:recursion}
\end{align}
where $\ddot{\mathbf{M}}_{t} = \tilde{\bm{\lagparam}}_t \tilde{\mathbf{M}}_{t - 1}\tilde{\bm{\lagparam}}_t^\top\nonumber$.
\end{theorem}
The proof is given in \ref{proof:recursion}. For response models with a linear mean-variance relationship it is possible to solve the system \eqref{eq:recursion} of recursive equations analytically; see \ref{proof:linear} for the proof. This applies to the Poisson, Gaussian, Laplace and uniform models from Table \ref{tab:qrs}.

\begin{theorem} 
Provided that an $\MPAR(G, Q, 0)$ process is second-order periodically stationary and $a_{gt} = 0$ for all $g$ and $t$ the unconditional second moments can be obtained as
$$
\vect(\tilde{\mathbf{M}}_{t})_{-1} = \left\{\mathbf{I}_{(QG + 1)^2 - 1} - (\tilde{\bm{\Phi}}_t \otimes \tilde{\bm{\Phi}}_t)_{-1, -1}\right\}^{-1} \left\{(\tilde{\bm{\Phi}}_t \otimes \tilde{\bm{\Phi}}_t)_{-1, 1} + \vect(\mathbf{\Xi}_t)_{-1}\right\}
$$
with $\mathbf{\Xi}_t = \diago(\tilde{\mathbf{b}}_t \circ \tilde{\bm{\mu}}_{t} + \tilde{\mathbf{c}}_t) + \sum_{i = 1}^{L - 1} \tilde{\bm{\phi}}_t \cdots\tilde{\bm{\phi}}_{t - i + 1}\diago(\tilde{\mathbf{b}}_{t - i} \circ \tilde{\bm{\mu}}_{t - i} + \tilde{\mathbf{c}}_{t - i})\tilde{\bm{\phi}}_{t - i + 1}^\top\cdots \tilde{\bm{\phi}}_t^\top$. The $\tilde{\bm{\mu}}_{gt}$ entering here are computed as in Remark \ref{remark_mu}. The index ``$-1$'' means that the first element/row/column is omitted while "1" means that only the first element/row/column is used. The operator $\circ$ denotes the entrywise product. 
\label{theorem:linear}
\end{theorem}

\begin{remark}
Note that Theorem \ref{theorem:linear} also applies to univariate models with higher lags. Notably if we include lagged $\lambda_t$ as described in Section \ref{sec:R_larger_0} our results generalize those by \cite{Bentarzi2017} on the periodic Poisson model. It seems that even for time-homogeneous models the corresponding expressions for unconditional moments are not yet available in the literature.\label{remark_linear}
\end{remark}

\noindent For the negative binomial and gamma models where generally $a_{gt} > 0$ we resort to solving the system \eqref{eq:recursion} numerically. Since $\tilde{\mathbf{M}}_{t} = \tilde{\mathbf{M}}_{t + L}$ it is sufficient to compute $\tilde{\mathbf{M}}_1, \dots \tilde{\mathbf{M}}_L$.

\begin{algo} Iterative computation of $\tilde{\mathbf{M}}_1, \dots, \tilde{\mathbf{M}}_L$
\label{algo:algo}
\begin{tabbing}
\enspace Set $\tilde{\mathbf{M}}_{0} = \tilde{\bm{\mu}}_L\tilde{\bm{\mu}}_L^\top$ with $\tilde{\bm{\mu}}_L$ calculated as in Remark \ref{remark_mu}.\\
\enspace Repeat until convergence or a maximum number of iterations:\\
\qquad For $t = 1, \dots, L$: (Re-) calculate $\tilde{\mathbf{M}}_{t}$ from $\tilde{\mathbf{M}}_{t- 1}$ using equation \eqref{eq:recursion} from Theorem \ref{theorem:recursion}.\\
\qquad Set $\tilde{\mathbf{M}}_{0} = \tilde{\mathbf{M}}_{L}$ for the next iteration. If successfully converged: return $\tilde{\mathbf{M}}_{1}, \dots, \tilde{\mathbf{M}}_L$.
\end{tabbing}
\end{algo}

\noindent Algorithm \ref{algo:algo} can be seen as a variant of a stationary iterative method \citep{Barrett1994} and extends an algorithm from \citeauthor{Held2017} (\citeyear{Held2017}, Appendix A) where we only considered conditional moments and limited ourselves to the negative binomial case and first order lags.

The $\tilde{\mathbf{M}}_t$ allow one to obtain unconditional means $\mu_{gt}$, variances $\sigma^2_{gt}$ and auto/cross-covariances $\gamma_{g'gt}(d)$ up to a time shift of $d = Q - 1$. To address the case $d \geq Q$ we can use a second recursive relationship
\begin{equation}
\E(\mathbf{Y}_t\mathbf{Y}_{t - k}^\top) = \mathbf{W}_t\E(\tilde{\mathbf{Y}}_{t - 1}\mathbf{Y}^\top_{t - k}) \label{eq:fill_offdiags}
\end{equation}
where $\mathbf{W}_{t} = (\bm{\nu_t}\ \ \ \bm{\phi}_{Qt} \ \ \ \cdots \ \ \ \bm{\phi}_{1t}) \in \mathbb{R}_{\geq 0}^{G\times(1 + QG)}$, i.e.\ it consists of the last $G$ rows of $\tilde{\bm{\phi}}_t$; see proof of Theorem \ref{theorem:recursion}, \ref{proof:recursion}.

\subsection{Extension to the case $R > 0$}
\label{sec:R_larger_0}

Theorems \ref{theorem:recursion} and \ref{theorem:linear} and Algorithm \ref{algo:algo} in fact do not require the distributional assumptions for units $g = 1, \dots G$ to be the same as long as the $a_{gt}, b_{gt}, c_{gt}$ are specified accordingly. This can be used to address models with the more general mean structure given in \eqref{eq:model_equation_matrix} where, without loss of generality, we set $R = Q$. We now define
$$
\bm{Z}_t = \left(\begin{matrix}\bm{Y}_t\\ \bm{\lambda}_t \end{matrix}\right), \ \ \ \
\bm{\nu}^Z_t = \left(\begin{matrix}\bm{\nu}_t\\ \bm{\nu}_t \end{matrix}\right),  \ \ \ \
\bm{\phi}^Z_{qt} = \left(\begin{matrix}\bm{\phi}_{qt} & \bm{\kappa}_{qt}\\ \bm{\phi}_{qt} & \bm{\kappa}_{qt}\end{matrix}\right).
$$
The process $\{\mathbf{Z}_t; t\in \mathbb{N}\}$ then follows an $\MPAR(2G, Q, 0)$ model with mixed distributional assumptions and parameters $\bm{\nu}^Z_t$ and $\bm{\phi}^Z_{qt}$ to govern the mean structure. The dispersion parameters are given by $\mathbf{a}^Z_t = (\mathbf{a}_t^\top, \bm{0}_G^{\top})^\top, \mathbf{b}^Z_t = (\mathbf{b}_t^\top, \bm{0}_G^{\top})^\top$ and $\mathbf{c}^Z_t = (\mathbf{c}_t^\top, \bm{0}_G^{\top})^\top$ where the zeros reflect the fact that the $\bm{\lambda}_t$ follow deterministically from $\mathcal{F}_{t - 1}$. Algorithm \ref{algo:algo} or Theorem \ref{theorem:linear} can then be applied using these parameters.  

\section{Application to time series of infectious disease counts}
\label{sec:application}

\subsection{Norovirus incidence in Bremen and Lower Saxony}

\begin{figure}[htb]
\begin{knitrout}
\definecolor{shadecolor}{rgb}{0.969, 0.969, 0.969}\color{fgcolor}
\includegraphics[width=\maxwidth]{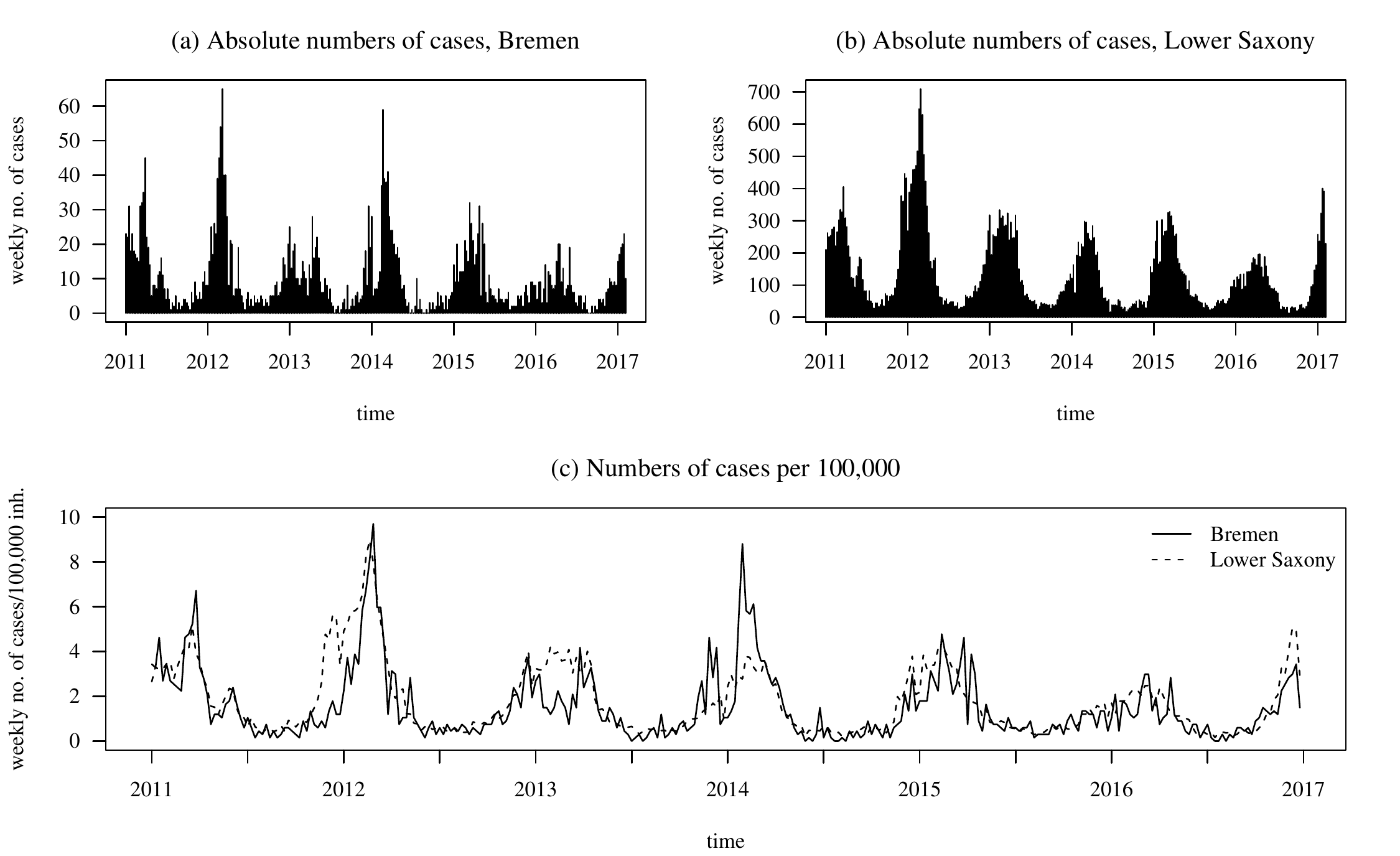} 

\end{knitrout}
\caption{Weekly case numbers of norovirus gastroenteritis in the German states of Bremen and Lower Saxony. Panels (a) and (b): absolute numbers. The scales in both plots differ as the population of  Lower Saxony is larger than that of Bremen. Panel (c): numbers of cases per 100,000 inhabitants, overlaid for the two states.}
\label{fig:incidence}
\end{figure}

Here we consider a bivariate time series of weekly case counts of norovirus gastroenteritis in Bremen ($B$), the smallest of Germany's 16 states (670,000 inhabitants), and the surrounding state of Lower Saxony ($L$; 7.9 million inhabitants). The data are freely available from the Robert Koch Institute at \url{https://survstat.rki.de} and cover weeks 2011/01 through 2016/52; the query was made on 30 May 2017. The mean yearly number of cases is 538.2 in Bremen (80.3/100,000 inhabitants) and 7362.5 in Lower Saxony (93.2/100.000 inhabitants). Norovirus gastroenteritis is one of the most common forms of food-borne gastroenteritis and has a mean generation time of 3--4 days. It shows very consistent seasonal patterns \citep{Rohayem2009}, making it a good showcase for periodic modelling. The data are shown in Fig. \ref{fig:incidence}.

\subsection{The endemic--epidemic modelling framework}
\label{subsec:hhh4}

The endemic--epidemic or $\hhh$ model is a multivariate time series model for infectious disease surveillance counts. It was introduced by \citet{Held2005} and is a special case of the class discussed here, specifically an $\MPAR(G, 1, 0)$ model with a Poisson or negative binomial conditional response model (see \citealt{Hoehle2016} for an overview). While the Poisson version is theoretically appealing as it can be motivated from discrete-time susceptible-infected-recovered (\textsc{sir}) models (see \citealt{Wakefield2017} on this), the more flexible negative binomial distribution usually offers a better fit in practice. For different types of stratification parsimonious parametrizations of the mean structure exist, including the use of power laws to describe spatial spread \citep{Meyer2014, Geilhufe2014} and social contact matrices for spread across age groups \citep{Meyer2017a}. Recent applications include studies on dengue in 21 districts of the Chinese province of Guangdong \citep{Cheng2016} and malaria and cutaneous leishmaniasis infection in 20 provinces of Afghanistan \citep{Adegboye2017}.

An assumption which is crucial for the scalability to higher dimensions is the conditional independence assumption $Y_{gt} \bot Y_{g't} \given \mathcal{F}_{t - 1}$ (eq. \ref{eq:dist_ass}). As mentioned in Section \ref{MPAR}, this is not to be confused with marginal independence of the different time series. In the context of multivariate infectious disease time series this assumption is common (see e.g.\ \citealt{Xia2004}, \citealt{Wang2011}) and justifiable if the generation time of the disease, i.e.\ the time span between symptoms in one infected person and those who get infected by this person is approximately equal to the reporting interval. Arguably, the assumption can be problematic when unobserved short term meteorological variations or other extrinsic drivers impact disease spread. It is thus recommended to check residual auto- and cross-correlations (see Section \ref{sec:results}) to see whether it is appropriate for the data at hand. Another assumption which deserves some consideration is that of non-negative parameters. As mentioned before, this forces all auto- and cross-correlations to be non-negative, too. While this may be limiting in other contexts, it is a natural feature of infectious disease spread.

Here we extend the framework to allow for higher order lags and set $Q = 5$. For the bivariate time series $Y_{gt}, g\in \{B, L\}, t=1, \dots, 312$ of weekly norovirus incidence in Bremen and Lower Saxony we specify the following distributed lag model:
\begin{equation}
Y_{gt} \sim \NegBin(\lambda_{gt}, \psi_g); \ \ \
\lambda_{gt} = \nu_{gt} + \sum_{g' \in \{B, L\}} \phi_{g'gt} \left(\sum_{q = 1}^5\lfloor w_q\rfloor Y_{g', t - q}\right)\label{eq:cond_means_noro}
\end{equation}
We reduce the number of parameters by specifying
\begin{align}
\log(\nu_{gt}) & = \log(e_g) + \alpha_g^{(\nu)} + \gamma_g^{(\nu)} \sin(\omega t) + \delta_g^{(\nu)} \cos(\omega t) \label{eq:nu}\\
\log(\phi_{ggt}) & =  \alpha_{g}^{(\phi)}+ \beta_{g}^{(\phi)}x_t + \gamma_{g}^{(\phi)} \sin(\omega t) + \delta_{g}^{(\phi)} \cos(\omega t)\label{eq:lags_own}\\
\log(\phi_{g'gt}) & = \log(e_g) + \alpha_{\times}^{(\phi)}+ \gamma_{\times}^{(\phi)} \sin(\omega t) + \delta_{\times}^{(\phi)} \cos(\omega t) \text{ \ \ \ \ for } g' \neq g \label{eq:cross_lags} \\
w_q & = p(1 - p)^{q - 1}; \lfloor w_q \rfloor  = w_q/\left(\sum_{q' = 1}^5 w_{q'}\right)\label{eq:hhh4_weights}
\end{align}
for $g, g' \in \{B, L\}$. The different components entering here are defined and interpreted as follows. The $\nu_{gt}$ represent the \textit{endemic} component and subsume climatic and socio-demographic factors as well as the population sizes $e_g$ which enter as offsets in \eqref{eq:nu}. Sine-cosine waves with frequencies $\omega = 2\pi/52$ are included to model yearly seasonality in weekly data \citep{Diggle1990, Held2012}. The \textit{epidemic} component $\sum_{g' \in \{B, L\}} \phi_{g'gt} \left(\sum_{q = 1}^5\lfloor w_q\rfloor Y_{g', t - q}\right)$  describes how incidence in a region $g$ is linked to previous incidence in regions $ g' = B, L$. Here we distinguish between the coefficients $\phi_{LLt}, \phi_{BBt}$ for lagged incidence from the same region, equation \eqref{eq:lags_own}, and coefficients $\phi_{LBt}, \phi_{BLt}$ for lagged incidence in the respective other region, equation \eqref{eq:cross_lags}. The parameters $\phi_{LLt}$ and $\phi_{BBt}$ are expected to play an important role, so we model them flexibly with unit-specific parameters, sine-cosine waves and an indicator $x_t$ for calendar weeks 1 and 52. The latter reflects reduced reporting during this period. As $\phi_{BLt}$ and $\phi_{LBt}$ are more difficult to identify we model them jointly using the parameters $\alpha^{(\phi)}_\times, \gamma^{(\phi)}_\times, \delta^{(\phi)}_\times$ and including offsets $\log(e_g)$ in \eqref{eq:cross_lags} to account for different population sizes. A similar strategy is used in \cite{Held2017}. Log-linear modelling of all coefficients is chosen in order to ensure positivity.

In \eqref{eq:cond_means_noro} and \eqref{eq:hhh4_weights} we introduce normalized lag weights $\lfloor w_q\rfloor$ in order to relax the usual assumption of $Q = 1$; in the more general notation from \eqref{eq:model_equation_matrix} we would denote $\phi_{g'gt}\lfloor w_q\rfloor$ by $(\phi_{qt})_{g'g}$. The geometric structure, i.e.\ an exponential decay steered by just one parameter $p$, is chosen for technical simplicity, but we parallel much of the compartmental modelling literature where exponentially distributed generation times are assumed. For simplicity we truncate the lags at $Q = 5$ weeks.

Sine-cosine waves with higher frequencies could easily be added to equations \eqref{eq:nu}--\eqref{eq:cross_lags}, the Akaike information criterion $\AIC$ serves to determine whether this is necessary. The endemic-epidemic model also allows for inclusion of covariates and time trends into the different components, but by doing so we would obviously leave our class of periodic autoregressive models. The overdispersion parameters $\psi_g$ are region-specific, but time constant in order to ensure identifiability.

Likelihood inference for the endemic--epidemic model with lag one is implemented in the \texttt{R} package \texttt{surveillance} \citep{Meyer2017} which is available on CRAN. Details can be found in the appendices of \cite{Paul2008} and \cite{Paul2011}; we extended this procedure to fit models with distributed lags. The parameter $p$ to steer the lag weights is estimated via a profile likelihood approach. 
An alternative would be full Bayesian inference, e.g.\ using a package like \texttt{JAGS}. However, as prior choices are not obvious we prefer the profile likelihood approach.

\subsection{Results}
\label{sec:results}

\begin{figure}[htb] 
\begin{knitrout}
\definecolor{shadecolor}{rgb}{0.969, 0.969, 0.969}\color{fgcolor}
\includegraphics[width=\maxwidth]{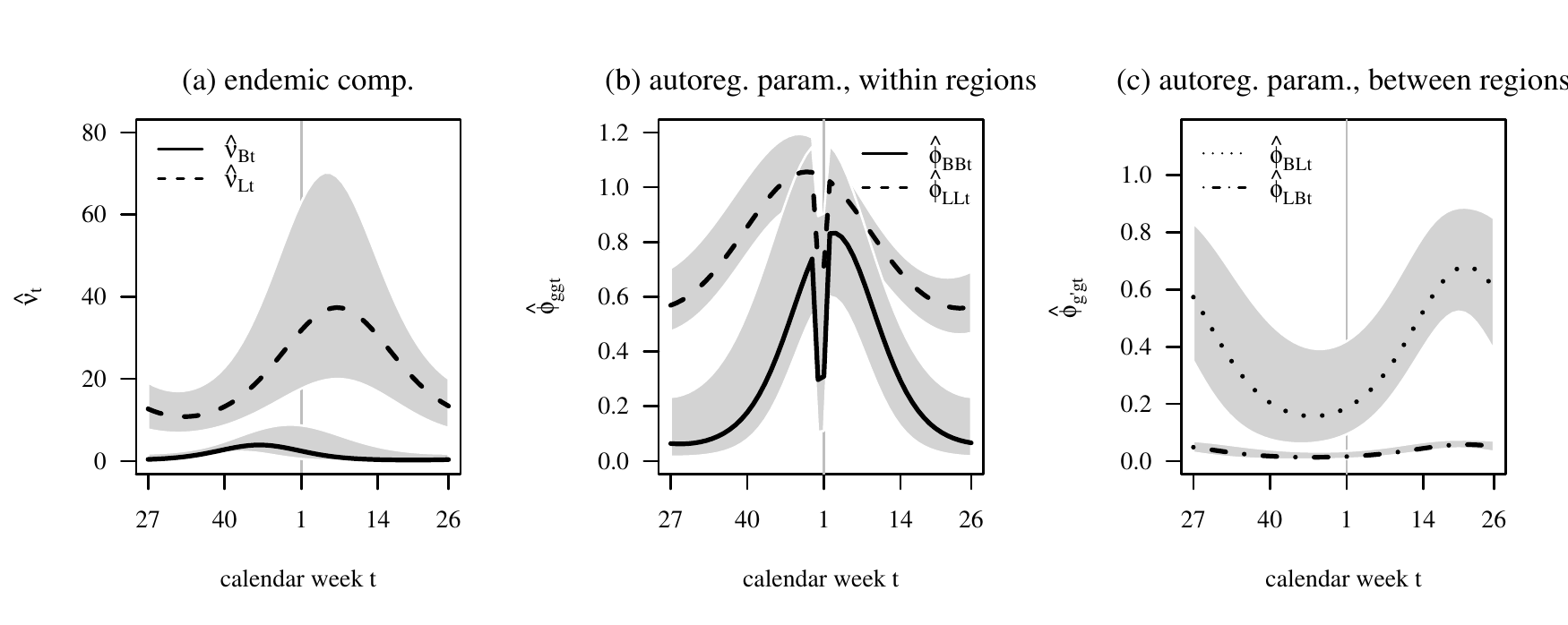} 

\end{knitrout}
\caption{Estimates for periodically varying parameters characterizing the mean structure in model \eqref{eq:cond_means_noro}--\eqref{eq:hhh4_weights} along with week-wise 95\% confidence intervals. Values are displayed over the course of a norovirus season starting in calendar week 27.}
\label{fig:parameters}
\end{figure}

In the following we apply the described model to the norovirus data; code for fitting the model and reproducing selected figures can be found online at \url{github.com/jbracher/hhh4addon}. The parameter $p$ is estimated as $0.73$ which gives $\lfloor \hat{w}_1\rfloor = 0.73$, $\lfloor \hat{w}_2\rfloor = 0.20$, $\lfloor \hat{w}_3\rfloor = 0.05$ as the three first lag weights. These values seem plausible given the short mean generation time of norovirus. The overdispersion parameters are estimated as $\hat{\psi}_B = 10.0$ (95\% CI from 7.2 to 14.0) and $\hat{\psi}_L = 26.0$ (95\% CI from $21.1$ to $32.0$). Estimates of $\nu_{gt}$ and $\phi_{g'gt}$ over the course of a norovirus season are displayed with week-wise confidence bands in Fig. \ref{fig:parameters}.

Due to their complex interplay, these parameters are difficult to interpret directly. Figure \ref{fig:parameters}(c) seems to indicate that preceding incidence in Bremen plays an important role in explaining incidence in Lower Saxony, but not the other way round. However, as the case numbers in Lower Saxony are generally much higher than in Bremen, the exact opposite is true, as will become clear in the following.

\begin{figure}[htb]
\begin{knitrout}
\definecolor{shadecolor}{rgb}{0.969, 0.969, 0.969}\color{fgcolor}
\includegraphics[width=\maxwidth]{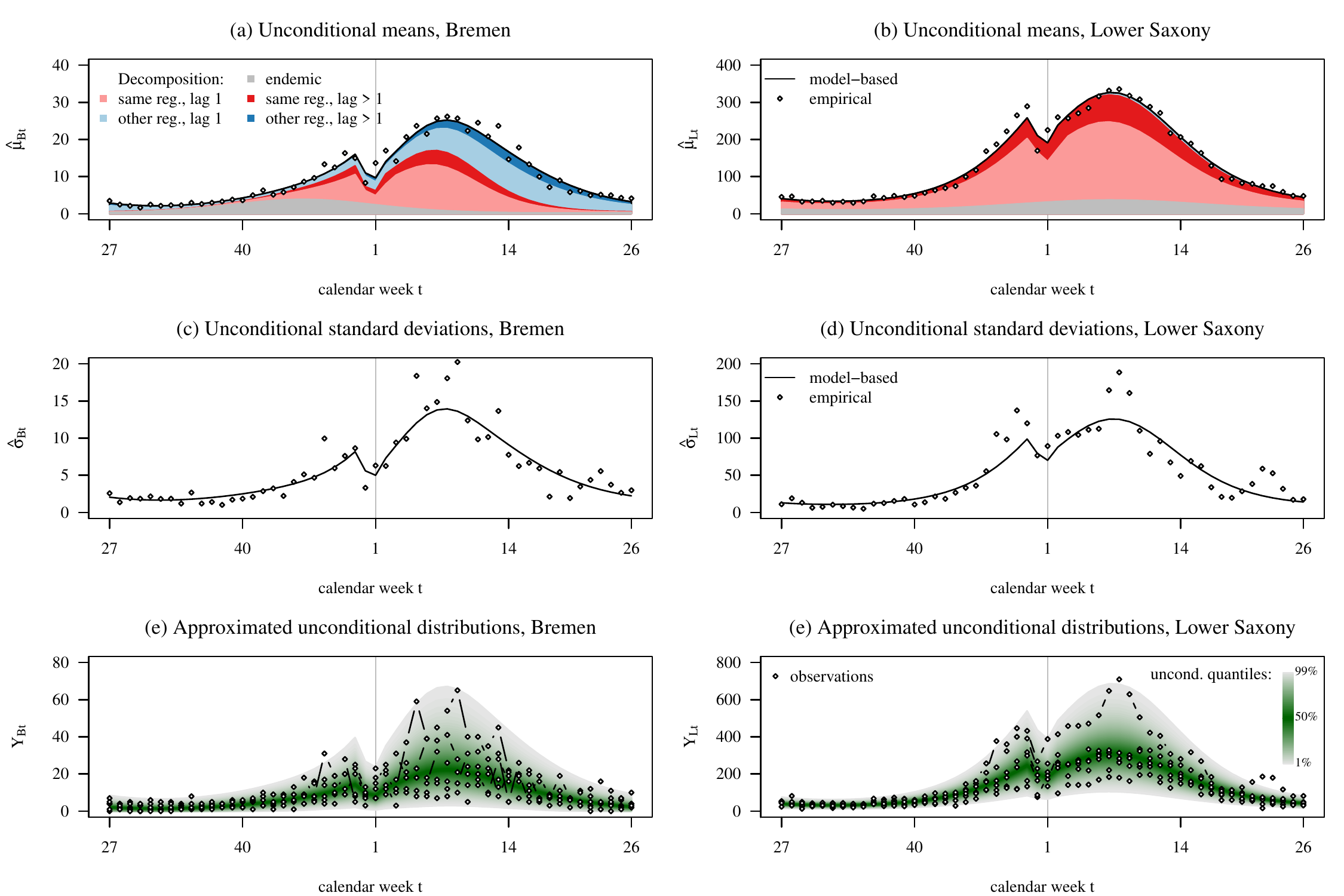} 

\end{knitrout}
\caption{Unconditional properties of the fitted model. Panels (a)--(d): Weekly unconditional means and standard deviations as implied by the fitted model (solid lines) compared to non-model-based empirical estimates (diamonds). Means are decomposed into contributions of the different components and lags. Panels (e) and (f): Negative binomial approximations of week-wise unconditional distributions and observations from the six years covered by our data.
}
\label{fig:stationary_moments}
\end{figure}

A visualization of the model fit is provided in Fig. \ref{fig:fit} in \ref{appendix:suppl_figures}. The model adapts well to the dynamics in the time series and the assumption of stable seasonality seems justified. The unconditional properties of the model are displayed in Fig. \ref{fig:stationary_moments}. The means $\hat{\mu}_{gt}$ in Panels (a) and (b) are decomposed into the different components and lags. The decomposition implies that over the course of the year, the model attributes 42\% of the incidence in Bremen to previous incidence in Lower Saxony, and 2.7\% in the reverse direction. The agreement with week-wise empirical estimates (means of the six available values per calendar week) is remarkable. The unconditional standard deviations $\hat{\sigma}_{gt}$ in Panels (c) and (d) show slightly less agreement, but this may be due to the poor stability of the week-wise empirical standard deviations. The bottom panels (e) and (f) show approximations of the week-wise unconditional distributions. These combine the quantities from the preceding plots via week-wise negative binomial approximations and provide an intuitive display of the variability in norovirus incidence curves. Simulation studies showed that these approximations usually work quite well. The observed curves are covered well by the distributions extracted from the model.

\begin{figure}[htb]
\begin{knitrout}
\definecolor{shadecolor}{rgb}{0.969, 0.969, 0.969}\color{fgcolor}
\includegraphics[width=\maxwidth]{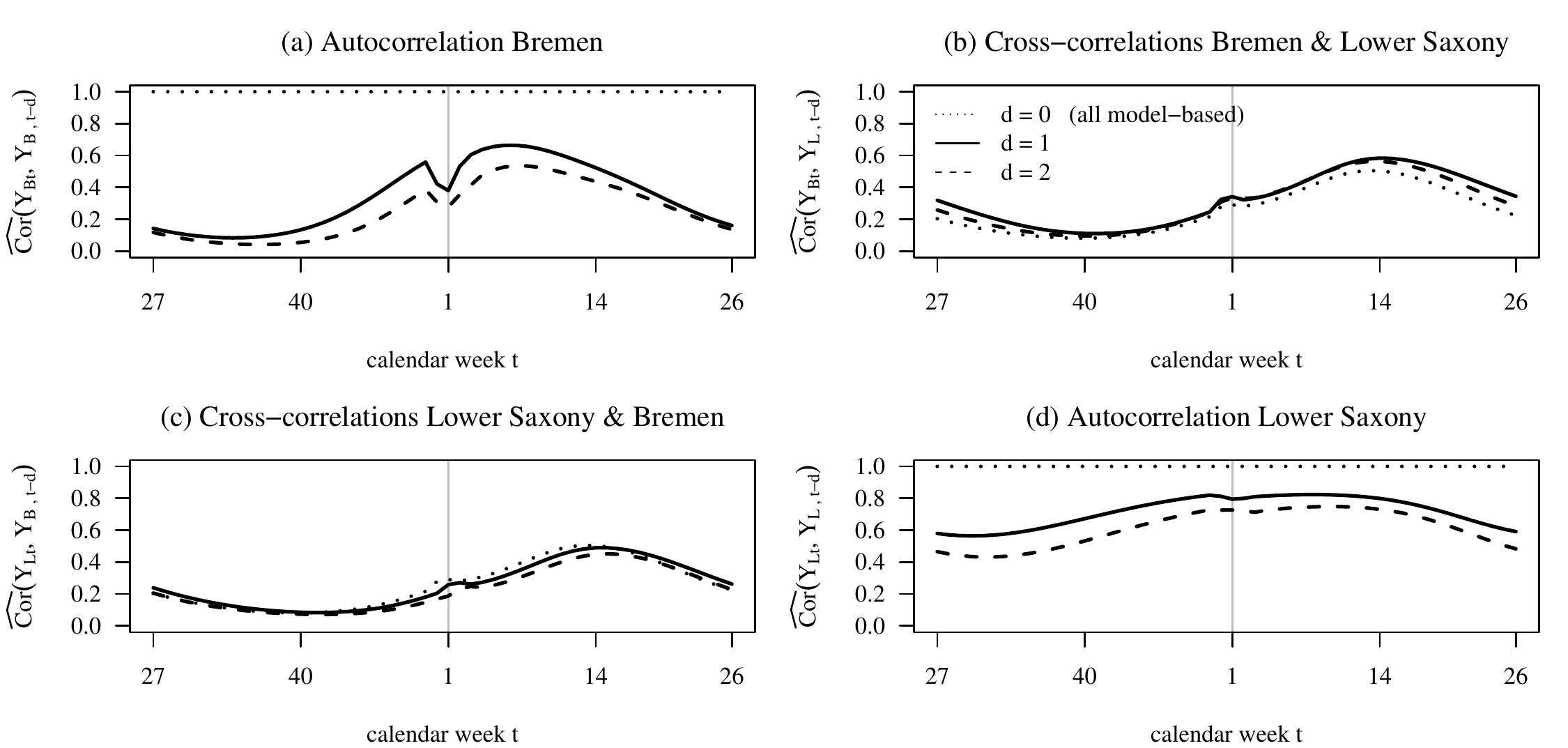} 

\end{knitrout}
\caption{Unconditional auto- and cross-correlation structure of the fitted bivariate model for lags $d = 0, 1, 2$: (a) $\Corr(Y_{Bt}, Y_{B, t - d})$, (b) $\Corr(Y_{Bt}, Y_{L, t - d})$, (c) $\Corr(Y_{Lt}, Y_{B, t - d})$ and (d) $\Corr(Y_{Bt}, Y_{B, t - d})$.
}
\label{fig:correlations}
\end{figure}

\begin{figure}[htb]
\begin{knitrout}
\definecolor{shadecolor}{rgb}{0.969, 0.969, 0.969}\color{fgcolor}
\includegraphics[width=\maxwidth]{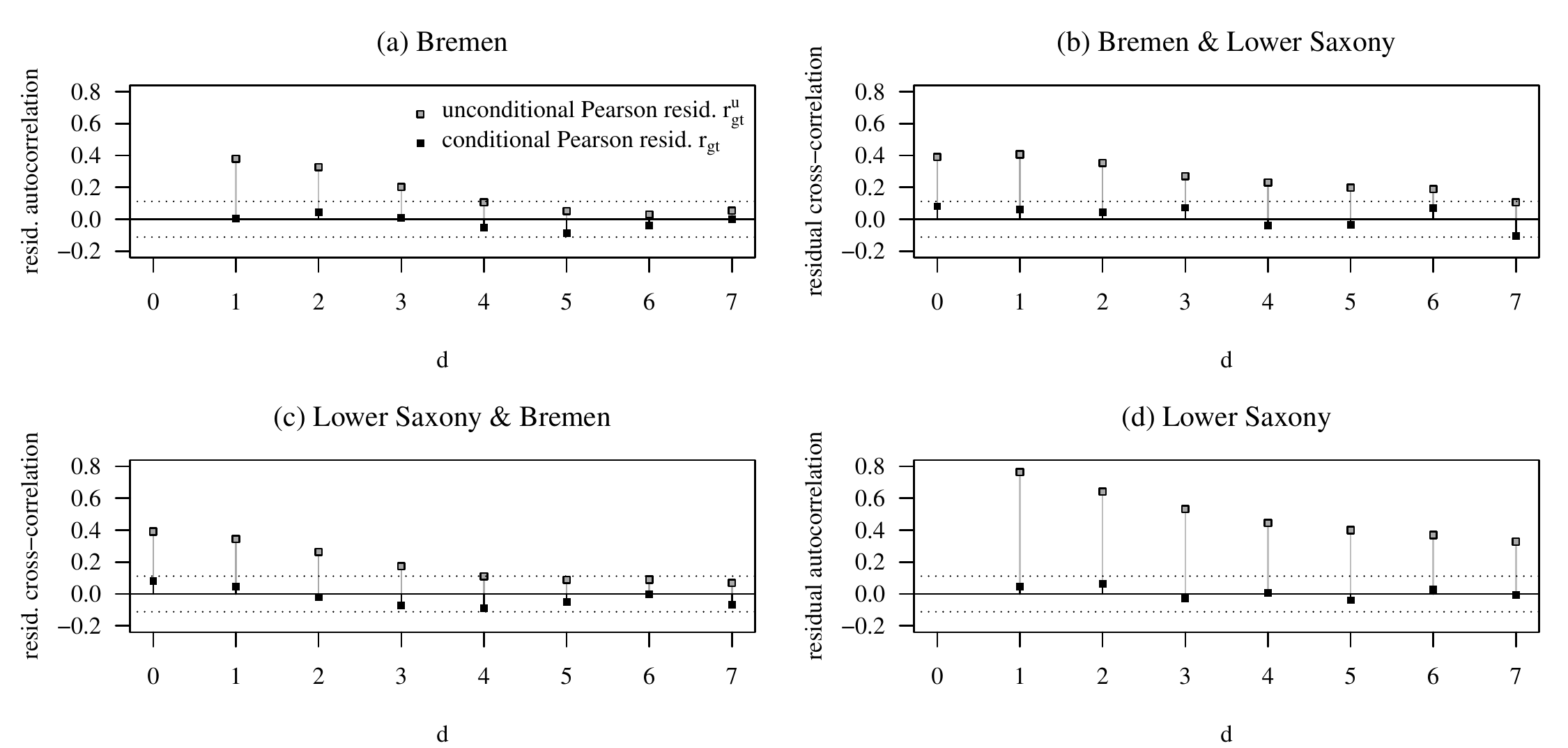} 

\end{knitrout}
\caption{Auto- and cross-correlation structure of the Pearson residuals $r_{gt}$ (black) and their unconditional counterparts $r_{gt}^{u}$ (grey), see equation \eqref{eq:Pearson}. (a) $\Corr(r_{Bt}, r_{B, t - d})$ and $\Corr(r^u_{Bt}, r^u_{B, t - d})$ (b) $\Corr(r_{Bt}, r_{L, t - d})$ and $\Corr(r^u_{Bt}, r^u_{L, t - d})$ (c) $\Corr(r_{Lt}, r_{B, t - d})$ and $\Corr (r^u_{Lt}, r^u_{B, t - d})$ (d) $\Corr(r_{Lt}, r_{L, t - d})$ and $\Corr(r^u_{Lt}, r^u_{L, t - d})$. The dotted 95\% significance line is at $\pm 1.96/307^{1/2} = \pm 0.11$ (where $6\cdot 52 - 5 = 307$ is the number of weeks for which fitted values are available when using five lags).
}
\label{fig:residuals}
\end{figure}

Next we explore the auto- and cross-correlation structure of the model. We display this in Fig. \ref{fig:correlations}; we omit empirical estimates here as they are very erratic. One can see that cross-correlation is generally stronger after the annual peak, i.e.\ the two time series are more closely intertwined. This is also reflected in the decomposition of the unconditional means $\mu_{Bt}$ for Bremen, see Fig. \ref{fig:stationary_moments}(a). To assess whether the correlation structure is captured adequately by the model we examine the residual correlation structure and compute conditional and unconditional Pearson residuals
\begin{equation}
r_{gt} = \frac{Y_{gt} - \hat{\lambda}_{gt}}{\sqrt{\hat{\psi}_g^{-1}\hat{\lambda}_{gt}^2 + \hat{\lambda}_{gt}}} \text{ \ \ \ and \ \ \ } r_{gt}^{u} = \frac{Y_{gt} - \hat{\mu}_{gt}}{\hat{\sigma}_{gt}},\label{eq:Pearson}
\end{equation}
respectively. While the $r_{gt}^{u}$ reflect the auto-correlation structure of the data, the $r_{gt}$ should not show any pattern. This is the case as is shown in Fig. \ref{fig:residuals}. In particular, the small cross-correlations of the conditional residuals $r_{gt}$ in Panels (b) and (c) suggest that the conditional independence assumption does not pose a problem here.


A comparison with simpler models shows that both distributed lags and the inclusion of lagged incidence from the respective other region are relevant. The $\AIC$ of the described model using both is 4446.9. The increase in AIC is 17.9 for a model containing only first lags and 26.5 for a model without coupling of the two regions. Omission of both features results in an increase of 51.2.

\section{Conclusion and outlook}
\label{sec:conclusion}

In this article we introduced a class of multivariate periodic autoregressive models with linear conditional means and applied a special case, the endemic--epidemic model, to  a dataset on norovirus gastroenteritis in Northern Germany. To this end we extended the model and the estimation procedure to allow for distributed lags. The estimated model parameters were not easily interpretable due to their complex interplay, especially as the considered geographical units differed profoundly in population size. Description of the model by its unconditional properties proved to be more informative and thus represents a useful tool for practical data analysis. In this article we showed a relatively simple bivariate real-data example. However, we also tested Algorithm \ref{algo:algo} on models for influenza incidence in $G = 140$ districts of Southern Germany \citep{Meyer2014} and norovirus incidence in six age groups and twelve districts of Berlin ($G = 72$; \citealp{Meyer2017}), and in each case it showed almost immediate convergence.

Insights from the present article can be of value in the future development of the endemic--epidemic model in various ways. They may be useful in embedding sophisticated models into existing algorithms for outbreak detection (e.g.\ \citealt{Noufaily2013}), thus extending these to the multivariate case. Alarm thresholds could then be defined as functions of the unit-specific and seasonally varying unconditional moments. Another potential use is in the further development of calibration tests along the lines of \cite{Held2017} and \cite{Wei2017}. We are also examining the implications with respect to under-reporting in surveillance systems. Under-reporting changes important stochastic properties of observed time series, for instance by attenuating autocorrelations, and can lead to biased parameter estimates if ignored. The theoretical results from this paper may help to address these difficulties.

Outside of the endemic--epidemic framework, the gamma model could be of use to analyse multivariate continuous outcomes as often observed in meteorology and hydrology. From a more theoretical point of view, possible extensions of the present model class could relax the conditional independence assumption $Y_{gt} \bot Y_{g't} \given \mathcal{F}_{t - 1}$ from equation \eqref{eq:dist_ass} as is done for instance in the time-homogeneous multivariate Poisson model by \cite{Heinen2007}. The approach described by \cite{Doukhan2017} may be a good starting point to this end.

\appendix

\section{Proofs}
\label{appendix}

In the following proofs we use again the notational convention $\prod_{i = j}^k x_i = 1$ for $j > k$ from Section \ref{subsec:analytical_results}.

\subsection{Proof of Theorem \ref{theorem:means}}
\label{proof:means}

Let $\lambda_0 < \infty$. Then using iterated expectations and model equation \eqref{eq:means_univ} we can write
\begin{align}
\E(Y_L\given \lambda_0) & = \E(\lambda_L\given \lambda_0) = \nu_L + (\phi_L + \kappa_L)\cdot\left\{\nu_{L - 1} + (\phi_{L - 1} + \kappa_{L - 1}) \cdot (\nu_{L - 2} + \dots)\right\}\nonumber\\
& = \underbrace{\sum_{i = 0}^{L - 1} \left\{\nu_{L - i} \ \ \prod_{l = 0}^{i-1} (\lagparam_{L - l} + \kappa_{L - l})\right\}}_{\text{denote this by } g} \ \ \ + \ \ \ \lambda_0 \underbrace{\prod_{m = 1}^L (\lagparam_{m} + \kappa_m)}_{\text{denote this by } f}  \label{eq:means1}
\end{align}
and due to the periodicity of the $\nu_t, \phi_t, \kappa_t$
\begin{equation*}
\E(Y_{kL}\given \lambda_0) = \sum_{i = 0}^{k - 1} g f^{i} \ \ \ + \ \ \ \lambda_0 f^{k}.
\end{equation*}
For $k \rightarrow \infty$ and $f = \prod_{m = 1}^L (\phi_m + \kappa_m) < 1$ we get $\lambda_0f^{k} = 0$; convergence of the geometric series then implies that the unconditional means exist and are $\mu_{kL} =  g/(1 - f)$. This proves equation \eqref{eq:theorem_means} for $t = kL$; the cases $t = kL + 1, kL + 2, \dots$ follow directly. 

\subsection{Proof of Theorem \ref{theorem:second_mom}}
\label{proof:second_mom}

We start by considering the unconditional variances of $\lambda_t$. Again let $\lambda_0 < \infty$ and consider
\begin{align*}
\Var(\lambda_t \given \lambda_0) = & \E\left\{\Var(\lambda_t \given \lambda_{t - 1}) \given \lambda_0\right\} + \Var\left\{\E(\lambda_t \given \lambda_{t - 1}) \given \lambda_0\right\}\\
= & \E\left\{\Var(\nu_t + \phi_t Y_{t - 1} + \kappa_t \lambda_{t - 1} \given \lambda_{t - 1}) \given \lambda_0\right\} + \Var\left\{\E(\lambda_t \given \lambda_{t - 1}) \given \lambda_0\right\}\\
= & \E\left\{\phi_t^2 \Var(Y_{t - 1} \given \lambda_{t - 1}) \given \lambda_0\right\} + \Var\left\{\nu_t + (\phi_t + \kappa_t)\lambda_{t - 1} \given \lambda_0\right\}\\
= & \E\left\{\phi_t^2 (a_{t - 1} \lambda_{t - 1}^2 + b_{t - 1}\lambda_{t - 1} + c_{t - 1}) \given \lambda_0\right\} + (\phi_t + \kappa_t)^2\Var(\lambda_{t - 1} \given \lambda_0)\\
%
%
%
= & \phi_t^2 \varfct_{t - 1}\left\{\E(\lambda_{t - 1} \given \lambda_0)\right\} + \left\{(\phi_t + \kappa_t)^2 + \phi_t^2a_{t - 1}\right\}\Var(\lambda_{t - 1} \given \lambda_0).
\end{align*}
Provided that the periodically stationary means exist $\phi_t^2 \varfct_{t - 1}\left\{\E(\lambda_{t - 1} \given \lambda_0)\right\}$ converges to \\ $\phi_t^2 \varfct_{t - 1}(\mu_{t - 1})$ such that for sufficiently large $t$ we can write
$$
\Var(\lambda_t \given \lambda_0) = \phi_t^2 \varfct_{t - 1}(\mu_{t - 1}) + \underbrace{\left\{(\phi_t + \kappa_t)^2 + \phi_t^2a_{t - 1}\right\}}_{\text{denote this by } h_t}\Var\left(\lambda_{t - 1} \given \lambda_0\right).
$$
Following the same arguments as in the proof of Theorem 1, \ref{proof:means}, we then get
$$
\Var(\lambda_{kL}) = \frac{\sum_{i = 0}^{L - 1} \phi^2_{L - i}\varfct_{L - i - 1}(\mu_{L - i - 1}) \prod_{l = 0}^{i - 1} h_{L - l}}{\prod_{m = 1}^{L} h_{m}},
$$
where $\prod_{m = 1}^{L} h_{m} = \prod_{m = 1}^L \left\{(\phi_m + \kappa_m)^2 + \phi_m^2a_{m - 1}\right\} < 1$ is required for convergence of the geometric series and thus existence of the variances. Now moving to $Y_{kL}$ we obtain
\begin{align}
\Var(Y_{kL}) = & \E\{\Var(Y_{kL} \given \lambda_{kL})\} + \Var\{\E(Y_{kL} \given \lambda_{kL})\}
\nonumber\\
= & \E(a_{kL} \lambda_{kL}^2 + b_{kL}\lambda_{kL} + c_{kL}) + \Var(\lambda_{kL})\nonumber \\
= & \varfct_{kL}(\mu_{kL}) + (a_{kL} + 1)\Var(\lambda_{kL})\label{eq:proof_var_y},
\end{align}
which proves equation \eqref{eq:theorem_vars} for $t = kL$. The cases $t = kL +1, kL +2, \dots$ follow immediately.
\subsection{Proof of Theorem \ref{theorem:covcor}}
\label{proof:covcor}

First note that
\begin{align*}
\Cov(Y_t, Y_{t - d}) & = \Cov\{\underbrace{\E(Y_t \given \mathcal{F}_{t - 1})}_{\lambda_t}, \underbrace{\E(Y_{t - d} \given \mathcal{F}_{t - 1})}_{Y_{t - d}}\} +  \E\{\underbrace{\Cov(Y_t, Y_{t - d} \given \mathcal{F}_{t - 1})}_{ = 0}\}\\
& = \Cov(\lambda_t, Y_{t - d})
\end{align*}
for $d \geq 1$ and then consider
$$
\Cov(\lambda_t, Y_{t - d}) = \Cov(\nu_t + \phi_t Y_{t - 1} + \kappa_t \lambda_{t - 1}, Y_{t - d})
 = \phi_t \Cov(Y_{t - 1}, Y_{t - d}) + \kappa_t\Cov(\lambda_{t - 1}, Y_{t - d}).
$$
While for $d = 1$ this becomes
\begin{align}
\Cov(\lambda_t, Y_{t - 1}) & = \phi_t\Var(Y_{t - 1}) + \kappa_t\Cov(Y_{t - 1}, \lambda_{t - 1}) = \phi_t\Var(Y_{t - 1}) + \kappa_t\Var(\lambda_{t - 1})\label{eq:case_d1}\nonumber\\
& = \phi_{t} \sigma^2_{t - 1} + \kappa_{t} \frac{\sigma^2_{t - 1} - \varfct_{t - 1}(\mu_{t - 1})}{a_{t - 1} + 1}
\end{align}
it simplifies to
$$
\Cov(\lambda_t, Y_{t - d}) = (\phi_t + \kappa_t)\Cov(Y_{t- 1}, Y_{t - d})\label{eq:case_d_larger_1}
$$
for $d > 1$. Equation \eqref{eq:theorem_cor} then follows by induction from \eqref{eq:case_d1} and \eqref{eq:case_d_larger_1}.

\subsection{Proof of Theorem \ref{theorem:recursion}}
\label{proof:recursion}

The matrix $\tilde{\bm{\phi}}_t$ is constructed such that $\tilde{\bm{\phi}}_t \tilde{\mathbf{Y}}_{t - 1} = (1, \mathbf{Y}_{t - Q + 1}^\top, \dots, \mathbf{Y}_{t - 1}^\top, \bm{\lambda}_t)^\top$; denote this vector by $\ddot{\mathbf{Y}}_t$. We now define $\ddot{\mathbf{M}}_t = \E(\ddot{\mathbf{Y}}_t\ddot{\mathbf{Y}}_t^\top)$ which we can express as
$$
\ddot{\mathbf{M}}_t = \E(\tilde{\bm{\phi}}_t \tilde{\mathbf{Y}}_{t - 1}\tilde{\mathbf{Y}}_{t - 1}^\top\tilde{\bm{\phi}}_t^\top) = \tilde{\bm{\phi}}_t \tilde{\mathbf{M}}_{t - 1} \tilde{\bm{\phi}}_t^\top
$$
where $\tilde{\mathbf{M}}_{t} = \E(\tilde{\mathbf{Y}}_{t}\tilde{\mathbf{Y}}_{t}^\top)$ and $\tilde{\mathbf{Y}}_{t} = (1, \mathbf{Y}_{t - Q + 1}^\top, \dots, \mathbf{Y}_{t}^\top)^\top$, as defined in Section \ref{subsec:algorithm_moments}. Already from their definition we see that $\ddot{\mathbf{M}}_t$ and $\tilde{\mathbf{M}}_t$ are largely the same. As $\E(\mathbf{Y}_t) = \E(\bm{\mu}_t)$,
$$
\E(Y_{gt}Y_{g't}) = \E\{\E(Y_{gt}Y_{g't} \given \mathcal{F}_{t - 1})\} = \E(\lambda_{gt}\lambda_{g't})
$$
for $g' \neq g$ and
$$
\E(\mathbf{Y}_{t - d}\mathbf{Y}_{t}) = \E\{\E(\mathbf{Y}_{t - d}\mathbf{Y}_{t}\given \mathcal{F}_{t - 1})\} = \E(\mathbf{Y}_{t - d}\bm{\lambda}_t),
$$
$\ddot{\mathbf{M}}_t$ and $\tilde{\mathbf{M}}_t$ are actually identical except for the last $G$ diagonal elements. These are $\E(\lambda_{1t}^2), \dots, \E(\lambda_{Gt}^2)$ in $\ddot{\mathbf{M}}_t$, but $\E(Y_{1t}^2), \dots, \E(Y_{Gt}^2)$ in $\tilde{\mathbf{M}}_t$. To obtain the latter from the former we can use
\begin{align*}
\E(Y_{gt}^2) & = \Var(Y_{gt}) + \E(Y_{gt})^2 = \E\left\{\Var(Y_{gt} \given \lambda_{gt})\right\} + \Var\left\{\E(Y_{gt} \given \lambda_{gt})\right\} + \E(Y_{gt})^2\\
& = \E(a_{gt}\lambda_{gt}^2 + b_{gt}\lambda_{gt} + c_{gt}) + \Var(\lambda_{gt}) + \E(Y_{gt})^2\\
& = a_{gt}\E(\lambda_{gt}^2) + b_{gt}\mu_{gt} + c_{gt} + \E(\lambda_{gt}^2) - \mu_{gt}^2 + \mu_{gt}^2\\
& = (a_{gt} + 1)\E(\lambda_{gt}^2) + b_{gt}\mu_{gt} + c_{gt}
\end{align*}
which corresponds to the correction given in \eqref{eq:recursion}. The leading zeros in $\tilde{\mathbf{a}}_t, \tilde{\mathbf{b}}_t, \tilde{\mathbf{c}}_t$ ensure that only the last $G$ elements are affected; $\mu_{gt}$ is taken from the first row of $\tilde{\mathbf{M}}_t$. This completes the recursive relationship given in Theorem \ref{theorem:recursion}.

Equation \eqref{eq:fill_offdiags} holds as $\mathbf{W}_t$ is constructed such that $\mathbf{W}_t\tilde{\mathbf{Y}}_{t - 1} = \bm{\lambda}_t$. We then get $\E(\mathbf{Y}_t\mathbf{Y}_{t - k}^\top) = \E(\bm{\lambda}_t\mathbf{Y}_{t - k}^\top) = \E(\mathbf{W}_t\tilde{\mathbf{Y}}_{t - 1}\mathbf{Y}^\top_{t - k})$.

\subsection{Proof of Theorem \ref{theorem:linear}}
\label{proof:linear}

If the conditional mean-variance relationship of an $\MPAR(G, Q, 0)$ model is linear we can obtain the periodically stationary first and second moments analytically (given that they exist). As we assume $a_{gt} = 0$ for all $g, t$, equation \eqref{eq:recursion} simplifies and applied iteratively allows us to write
\begin{align}
\tilde{\mathbf{M}}_t & = \tilde{\bm{\phi}}_t \tilde{\mathbf{M}}_{t - 1} \tilde{\bm{\phi}}_t^\top + \diago(\tilde{\mathbf{b}}_t \circ \tilde{\bm{\mu}}_{t} + \tilde{\mathbf{c}}_t) =  \dots \label{eq:quad_eq}\\
& = \tilde{\bm{\Phi}}_t \tilde{\mathbf{M}}_{t - L} \tilde{\bm{\Phi}}_t^\top + \underbrace{\diago(\tilde{\mathbf{b}}_t \circ \tilde{\bm{\mu}}_{t} + \tilde{\mathbf{c}}_t) + \sum_{i = 1}^{L - 1} \tilde{\bm{\phi}}_t \cdots\tilde{\bm{\phi}}_{t - i + 1}\diago(\tilde{\mathbf{b}}_{t - i} \circ \tilde{\bm{\mu}}_{t - i} + \tilde{\mathbf{c}}_{t - i})\tilde{\bm{\phi}}_{t - i + 1}^\top\cdots \tilde{\bm{\phi}}_t^\top}_{\text{denote this by } \mathbf{\Xi}_t}\nonumber
\end{align}
where the $\bm{\mu}_t$ are known from Remark \ref{remark_mu} and $\tilde{\bm{\Phi}}_t = \tilde{\bm{\lagparam}}_{t} \tilde{\bm{\lagparam}}_{t - 1} \cdots \tilde{\bm{\lagparam}}_{t - L + 1}$. Note that in \eqref{eq:recursion} the $\mu_{gt}$ are taken directly from the first row of $\tilde{\mathbf{M}}_t$. Using that $\tilde{\mathbf{M}}_t = \tilde{\mathbf{M}}_{t + L}$ we can vectorize and and re-order \eqref{eq:quad_eq} to
\begin{equation}
(\mathbf{I}_{(QG + 1)^2} - \tilde{\bm{\Phi}}_t \otimes \tilde{\bm{\Phi}}_t) \text{vect}(\tilde{\mathbf{M}}_{t}) = \text{vect}(\mathbf{\Xi}_t).\label{eq:Kronecker}
\end{equation}
This system is underdetermined because by construction the first line of $\mathbf{I}_{(QG + 1)^2} - \tilde{\bm{\Phi}}_t \otimes \tilde{\bm{\Phi}}_t$ is the zero vector $\mathbf{0}_{(QG + 1)^2}$ and $\vect(\bm{\Xi})_1 = 0$. However, this is made up for by the constraint $\text{vect}(\tilde{\mathbf{M}}_{t})_1 = 1$ which follows from the definition of $\tilde{\mathbf{M}}_{t}$ in Section \ref{subsec:algorithm_moments}. If we set $\text{vect}(\tilde{\mathbf{M}}_t)_1 = 1$ we can reduce the dimension of the quadratic forms in \eqref{eq:Kronecker} by one and handle the first column of $(\tilde{\bm{\Phi}}_t \otimes \tilde{\bm{\Phi}}_t)$ separately:
$$
\left\{\mathbf{I}_{(QG + 1)^2 - 1} - (\tilde{\bm{\Phi}}_t \otimes \tilde{\bm{\Phi}}_t)_{-1, -1}\right\} \text{vect}(\tilde{\mathbf{M}}_{t})_{-1} - (\tilde{\bm{\Phi}}_t \otimes \tilde{\bm{\Phi}}_t)_{-1, 1} = \text{vect}(\mathbf{\Xi}_t)_{-1}
$$
Here the index ``$-1$'' means that the first element/row/column is omitted while "1" means that only the first element/row/column is used. This proves Theorem \ref{theorem:linear} as it results in
$$
\vect(\tilde{\mathbf{M}}_{t})_{-1} = \left\{\mathbf{I}_{(QG + 1)^2 - 1} - (\tilde{\bm{\Phi}}_t \otimes \tilde{\bm{\Phi}}_t)_{-1, -1}\right\}^{-1} \left\{(\tilde{\bm{\Phi}}_t \otimes \tilde{\bm{\Phi}}_t)_{-1, 1} + \text{vect}(\mathbf{\Xi}_t)_{-1}\right\}.
$$

\section{Supplementary figures on data analysis}
\label{appendix:suppl_figures}
\begin{figure}[htb]
\begin{knitrout}
\definecolor{shadecolor}{rgb}{0.969, 0.969, 0.969}\color{fgcolor}
\includegraphics[width=\maxwidth]{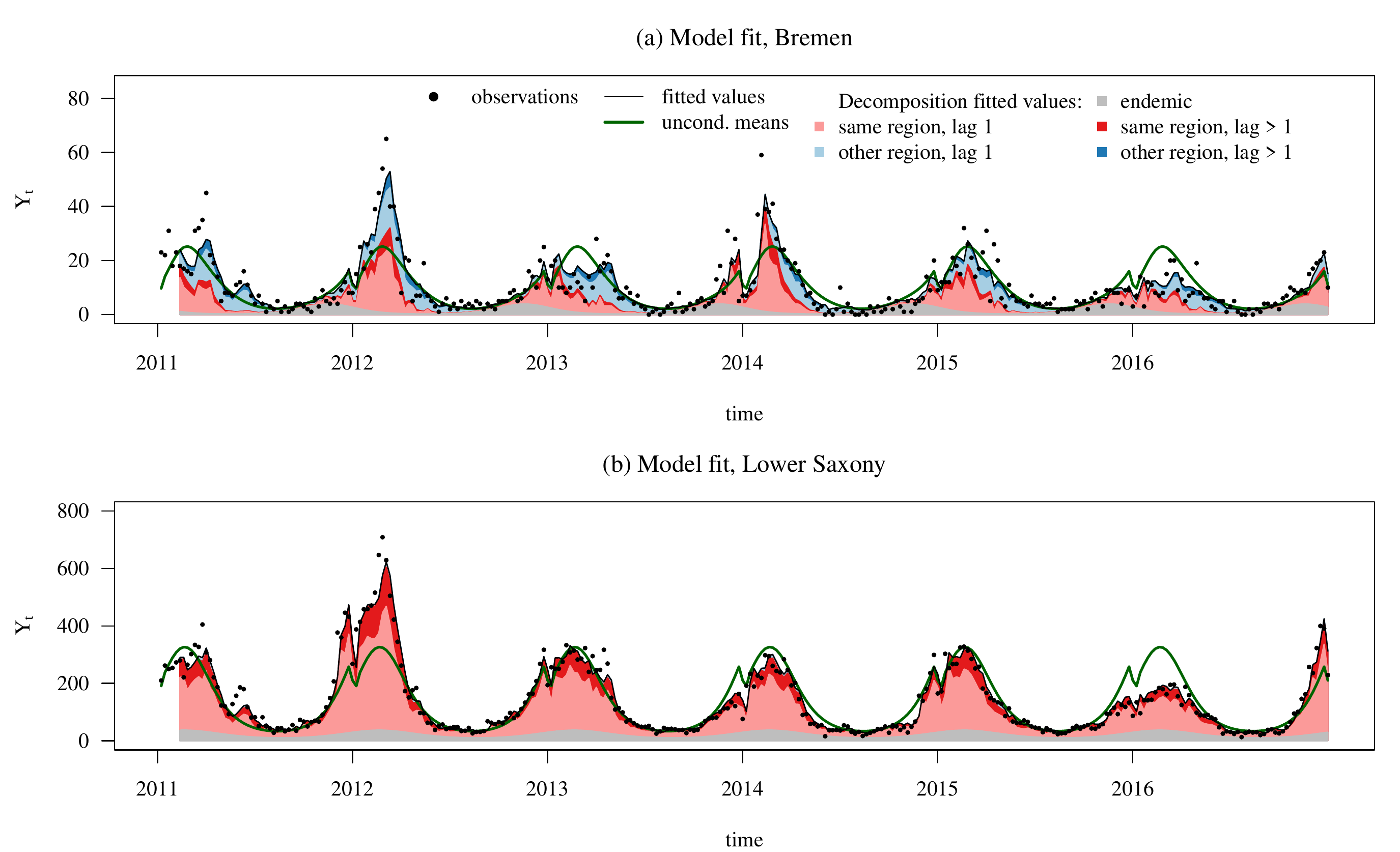} 

\end{knitrout}
\caption{Observations and fitted values for Bremen and Lower Saxony. Fitted values are decomposed into the contributions of the different components and lag orders (contributions of higher order lags in darker colour). Generally the models adapts well to the observed dynamics. While the amplitude of observed incidence varies considerably across the different seasons, the timing is well aligned with the unconditional moments of our model (added as solid green lines). The assumption of stable seasonality thus seems acceptable.}
\label{fig:fit}
\end{figure}

\section*{Acknowledgement}

We thank Kelly Reeve and Samuel Atwell for valuable input on the manuscript.

\bibliography{Bib_Arxive2.bib}
{\footnotesize
\bibliographystyle{apalike}}
\end{document}